\documentclass[12pt]{elsarticle}
\usepackage{amssymb}

\usepackage{caption}
\usepackage{subfigure}  
\usepackage{amsmath}
\usepackage{bm}% bold math
\usepackage{color}
\usepackage[colorlinks=True]{hyperref}

\begin{document}

\begin{frontmatter}
\title{A paradox of the Navier-Stokes turbulence
}

\author[label1,label2]{Shijie Qin}
\author[label2]{Kun Xu}
\author[label1,label3]{Shijun Liao \corref{cor1}}

\cortext[cor1]{sjliao@sjtu.edu.cn}

\address[label1]{State Key Laboratory of Ocean Engineering, Shanghai 200240, China}
\address[label2]{Department of Mathematics, Hong Kong University of Science and Technology, Hong Kong, China}
\address[label3]{School of Ocean and Civil Engineering, Shanghai Jiao Tong University, Shanghai 200240, China}
	
\begin{abstract}	

The Navier-Stokes (NS) equations as a turbulence model have been widely applied in lots of fields.  The NS equations contain such a fundamental assumption that all small physical/artificial disturbances could be neglected.  Is this assumption correct?  In this paper a two-dimensional Rayleigh-B\'{e}nard convection governed by the NS equations is  predicted by traditional direct numerical simulation (DNS) using double precision arithmetic and  a range of different time-steps.   It is found that the final flow type tends either to vortical flow or zonal flow, whose statistics are completely different.  Notably, these two flow types frequently alternate as the time-step  is reduced to a very small value, suggesting that the time-step corresponding to each turbulent flow type should be densely distributed.  Thus, stochastic numerical noise  exerts a huge influence on the final flow type and statistics of  numerically simulated NS turbulence because the time-step has a close relationship with numerical noise.   This clearly indicates that small disturbances have significant influences on the NS turbulence, which therefore should not be neglected.   This leads to a logical paradox  for  the NS turbulence, which is a great challenge for us, although a paradox often leads to some significant breakthroughs.    
\end{abstract}


\end{frontmatter}
			
\section{Introduction}

In 1890, Poincar\'{e}~\cite{Poincare1890} discovered that  the spatiotemporal trajectories of certain dynamic systems   exhibit sensitivity dependence on  the initial condition, later termed chaos, i.e.,  the famous ``butterfly-effect''  coined by Lorenz~\cite{Lorenz1963} in 1963 that led to the collapse of  pure determinism  as understood by the scientific community~\cite{Smale1967, Li-Yorke1975, May1976, Feigenbaum1978, Feigenbaum1979}.      
Many researchers~\cite{Deissler1986PoF, boffetta2017chaos, berera2018chaotic, Vassilicos2023JFM, Ge2025JFM, Ge2025PRF, Okamoto2025PRF}  have reported that  the spatiotemporal trajectories  of Navier-Stokes turbulence (i.e., turbulent flow governed by the Navier-Stokes  equations)   display sensitivity dependence on initial conditions   {and/or} arithmetic precision, corresponding to  so-called trajectory instability.   In other words,    Navier-Stokes (NS) turbulence is chaotic in that any tiny (natural/artificial) stochastic disturbances increase exponentially to the macroscopic level of spatiotemporal trajectories.  

 In 2006, Lorenz \cite{Lorenz2006Tellus}  further discovered that   numerical noise from truncation and round-off errors could exert a huge influence on the numerical simulations of chaotic systems   --  {\em not only} on their spatiotemporal trajectories {\em but also} even on their statistics.    Notably, Lorenz  \cite{Lorenz2006Tellus}   demonstrated that the  sign of the maximum Lyapunov exponent of a chaotic system  might alternate as the time-step decreases to a very small value.   It should be emphasized that   although initial conditions have physical meaning, numerical noise is inherently artificial and therefore artificially uncertain because of its strong dependence on numerical algorithms that might  be quite different even when invoked with the {\em same} governing equations and the {\em same} initial/boundary conditions.    

How large is the influence of numerical noise  on NS turbulence?   Could numerical noise lead to  inherent and statistical differences  in NS  turbulence?  These are  fundamentally important questions for   the turbulence modelling community  to consider.    

Why?  It should be emphasized that the NS equations neglect all small disturbances, no matter whether they are natural or artificial.   Note that NS equations were first proposed  by Navier~\cite{Navier1822} in 1822 and then modified by Stokes~\cite{Stokes1845} in 1845, many years  before  Poincar\'{e}~\cite{Poincare1890} proposed, for the first time,  the concept of sensitivity dependence on initial condition  in 1890, and more than one century before the chaos theory~\cite{Lorenz1963, Smale1967, Li-Yorke1975, May1976, Feigenbaum1978, Feigenbaum1979} was well established, and even many years before  the concept of turbulence was proposed by Reynolds~\cite{Reynolds1895} in 1895 and by Taylor~\cite{Taylor1935} in 1935.   So, it was very reasonable for Navier~\cite{Navier1822}  and Stokes~\cite{Stokes1845} to completely neglect the influence of all tiny disturbances in the Navier-Stokes equations for turbulent flows that are believed to be chaotic today.   However, if small artificial/natural disturbances indeed have significant influences on macro-scale property of the NS turbulence, a paradox does exist in logic, which should be a great challenge for us today.  

\begin{figure}[tb]
\vspace{-10pt} 
    \begin{center}
        \begin{tabular}{cc}
             \includegraphics[width=2.0in]{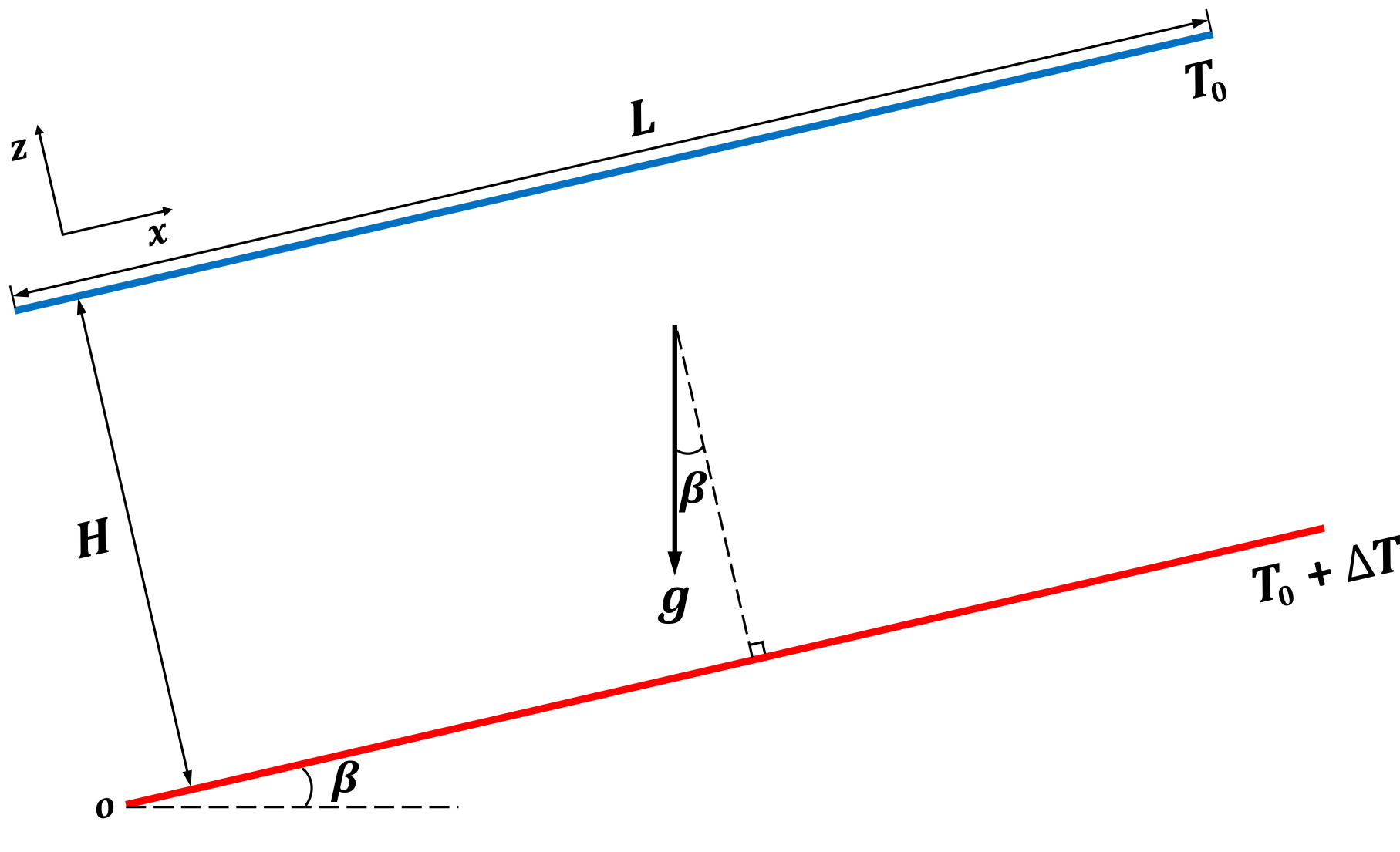}
        \end{tabular}
 \end{center}
 \vspace{-20pt} 
\caption{Schematic of 2D turbulent RBC  between two parallel plates tilted at an angle $\beta$. The fluid layer, between  the plates a distance $H$ apart,  is heated by the bottom (red) plate  and cooled by the top (blue) plate  such that there is a constant temperature difference $\Delta T>0$. Here $L$ is the width of  the computational domain, and gravitational acceleration $g$ is directed downward.}
\label{schematic_drawings}
\vspace{-10pt} 
\end{figure}

To answer these questions, this paper considers the case of two-dimensional  (2D)  turbulent Rayleigh-B\'{e}nard convection (RBC) \cite{Rayleigh1916On, wei2013viscous, shishkina2016thermal, subramanian2016spatio,  zwirner2018confined,  wang2020vibration}   { shown in FIG.~\ref{schematic_drawings}, within} a thin layer of fluid confined between two parallel plates separated by a distance $H$ and tilted at an angle $\beta$ (positive when the plates rotate anti-clockwise)  \cite{wang2018flow, wang2018multiple}.  The fluid gains heat from the bottom plate and is cooled by the top plate.  The $x$-axis is parallel with the plates, the $z$-axis is perpendicular to the  bottom inclined plate, $T_{0}$ and $T_{0}+\Delta T$ (with $\Delta T>0$) denote the temperatures of the top (blue) plate and bottom (red) plate,  and $g$ is the acceleration due to  gravity.

Taking the length scale $H$, velocity scale $\sqrt{g\alpha H\Delta T}$ (where $\alpha$ is the thermal expansion coefficient) and temperature scale $\Delta T$ as characteristic scales, the dimensionless Navier-Stokes equations with Boussinesq approximation \cite{saltzman1962finite} and tilting angle $\beta$ are as follows:
\begin{equation}
\begin{aligned}
\frac{\partial}{\partial t}\nabla^{2}\psi+\frac{\partial(\psi,\nabla^{2}\psi)}{\partial(x,z)}&-\frac{\partial\theta}{\partial x} \cos \beta
+\left(\frac{\partial\theta}{\partial z}-1\right) \sin \beta \\-&\sqrt{\frac{Pr}{Ra}}\nabla^{4}\psi=0,       \label{RB_psi}
\end{aligned}
\end{equation}
\begin{equation}
\frac{\partial\theta}{\partial t}+\frac{\partial(\psi,\theta)}{\partial(x,z)}-\frac{\partial\psi}{\partial x}-\frac{\nabla^{2}\theta}{\sqrt{PrRa}}=0,       \label{RB_theta}
\end{equation}
subject to the boundary conditions
\begin{equation}
\psi=\frac{\partial^{2}\psi}{\partial z^{2}}=\theta=0, \hspace{1.0cm}  \mbox{at $z=0$ and $1$}    \label{free-slip}
\end{equation}
on the upper and lower  plates respectively, and the periodic conditions
\begin{equation}
\psi(x,z,t)=\psi(x+\Gamma,z,t), \hspace{0.45cm} \theta(x,z,t)=\theta(x+\Gamma,z,t),    \label{periodic}
\end{equation}
where $\psi$ is  the stream function   {such that} $u=-\partial\psi/\partial z$ and $w=\partial\psi/\partial x$, in which $u$ and $w$ are   {the fluid} velocity components in the $x$- and $z$-directions, $t$ denotes time, $x\in[0,\Gamma]$ and $z\in[0,1]$ are position coordinates, $\Gamma = L/H$ is an aspect ratio, and $\theta$ denotes the temperature departure from a linear variation background (i.e., the dimensionless temperature is expressed as $T =1-z+\theta$), 
$\nabla^{2}$ denotes the Laplace operator with $\nabla^{4}=\nabla^{2}\nabla^{2}$, and 
$\partial(A,B)/\partial(x,z)=(\partial A/\partial x)(\partial B/\partial z)-(\partial B/\partial x)(\partial A/\partial z)$ is the Jacobi   {operator}.
The dimensionless Rayleigh number $Ra$ and Prandtl number $Pr$ are defined by
\begin{equation}
Ra=\frac{g\alpha H^{3}\Delta T}{\nu\kappa}, \hspace{1.0cm} Pr=\frac{\nu}{\kappa},       \label{Ra_Pr}
\end{equation}
where $\nu$  is fluid kinematic viscosity and $\kappa$ is thermal diffusivity.  The initial temperature and velocity fields are {\em randomly} generated as thermal fluctuations in the form of Gaussian white noise, with the temperature standard deviation $\sigma_T=10^{-10}$ and velocity standard deviation $\sigma_u=10^{-9}$.  It should be emphasized that {\em all} numerical simulations   {reported in} this paper have the {\em same} initial condition.  
We consider here the case   {for which} $\Gamma=2\sqrt{2}$ (which is large enough  to approximate the heat flux by the value for an infinite aspect ratio \cite{saltzman1962finite, curry1984order}), $Pr=6.8$ (corresponding to water at room temperature 20$^{\circ}$C) and $Ra=6.8\times10^{8}$ (corresponding to a turbulent flow).
	
\begin{figure}
    \begin{center}
        \begin{tabular}{cc}
             \includegraphics[width=3.5in]{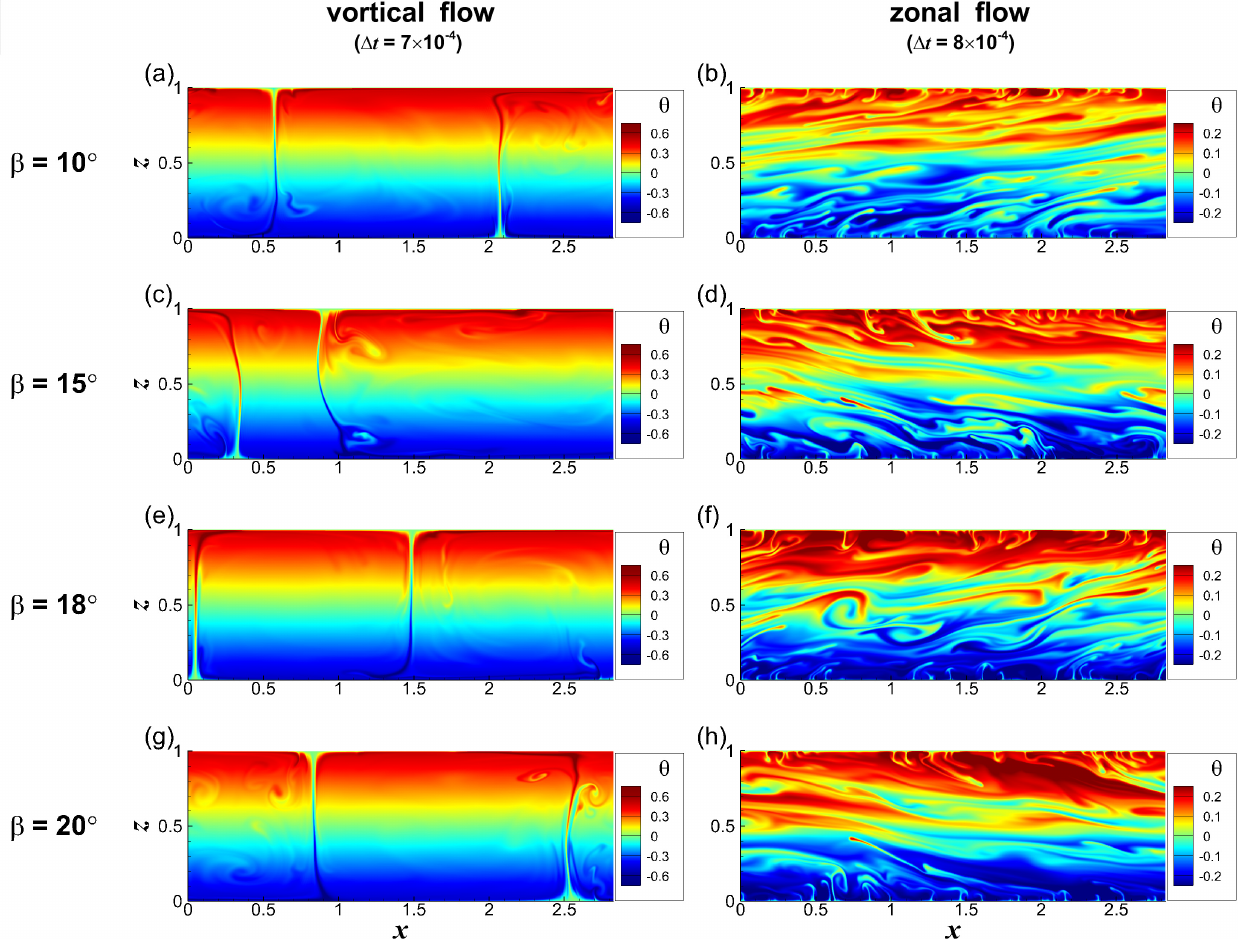}
        \end{tabular}
    \caption{$\theta$ (temperature departure from the linear variation background) fields at $t=500$ of the turbulent RBC tilted   {at angles} (a)-(b) $\beta=10^\circ$, (c)-(d) $\beta=15^\circ$, (e)-(f) $\beta=18^\circ$, and (g)-(h) $\beta=20^\circ$, given by DNS   {with time steps} $\triangle t=7\times10^{-4}$ (left, vortical flow) and $\triangle t=8\times10^{-4}$ (right, zonal flow).}     \label{Contour}
    \end{center}
\end{figure}
	
\begin{figure}
    \begin{center}
        \begin{tabular}{cc}
             \includegraphics[width=2.5in]{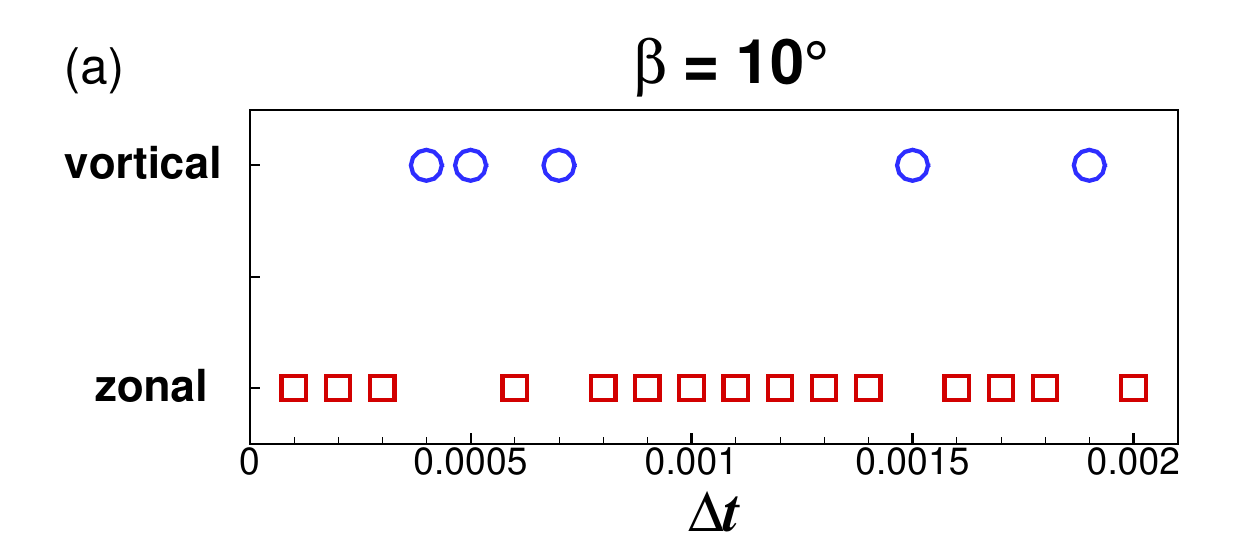}
             \includegraphics[width=2.5in]{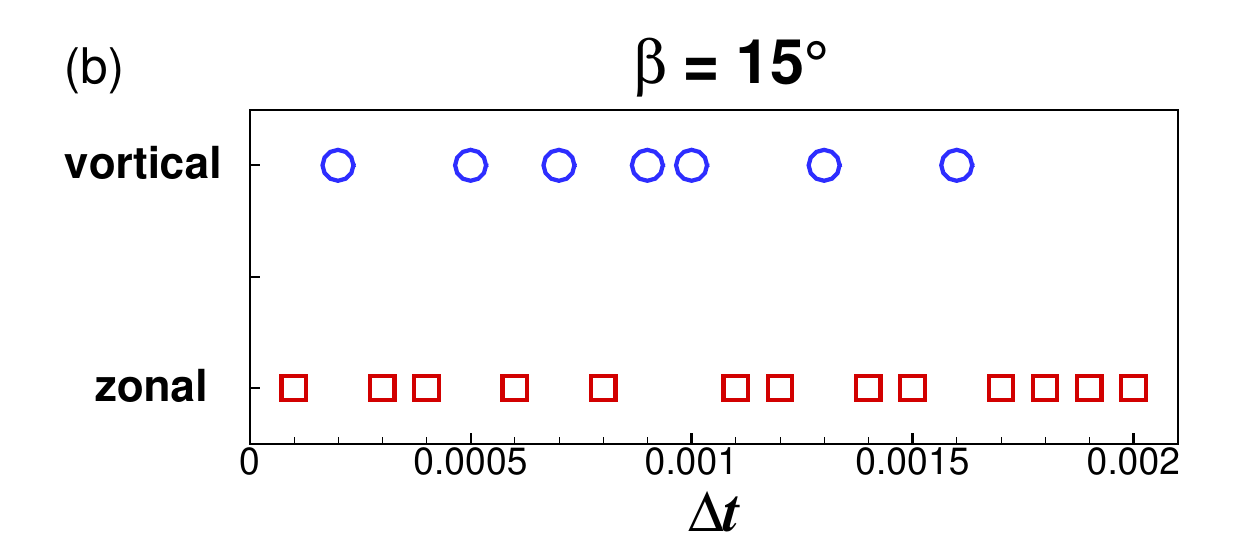}\\
             \includegraphics[width=2.5in]{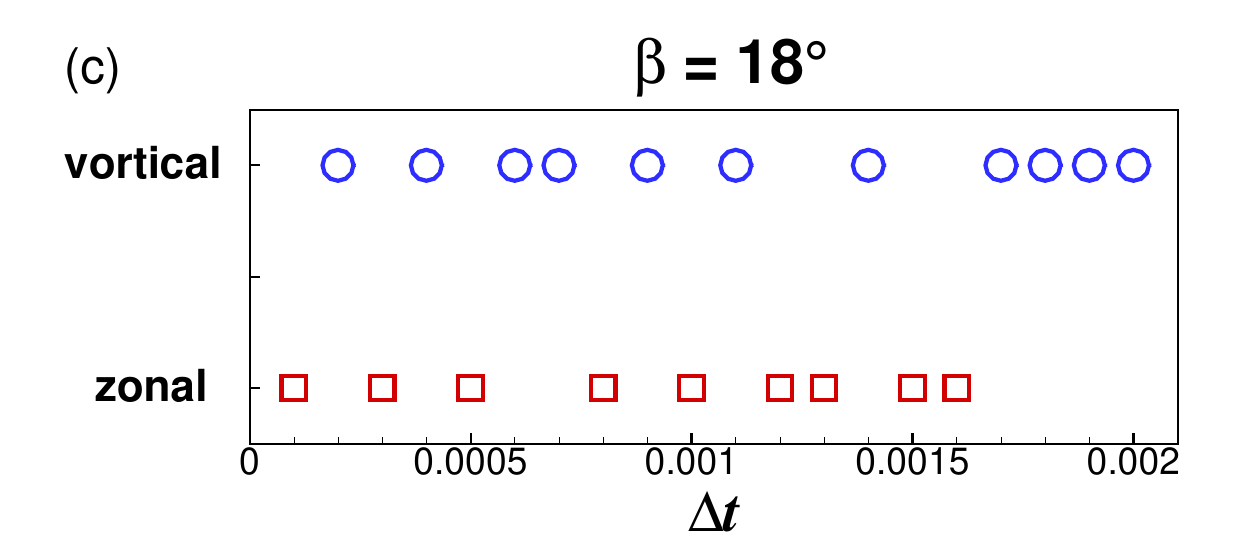}
             \includegraphics[width=2.5in]{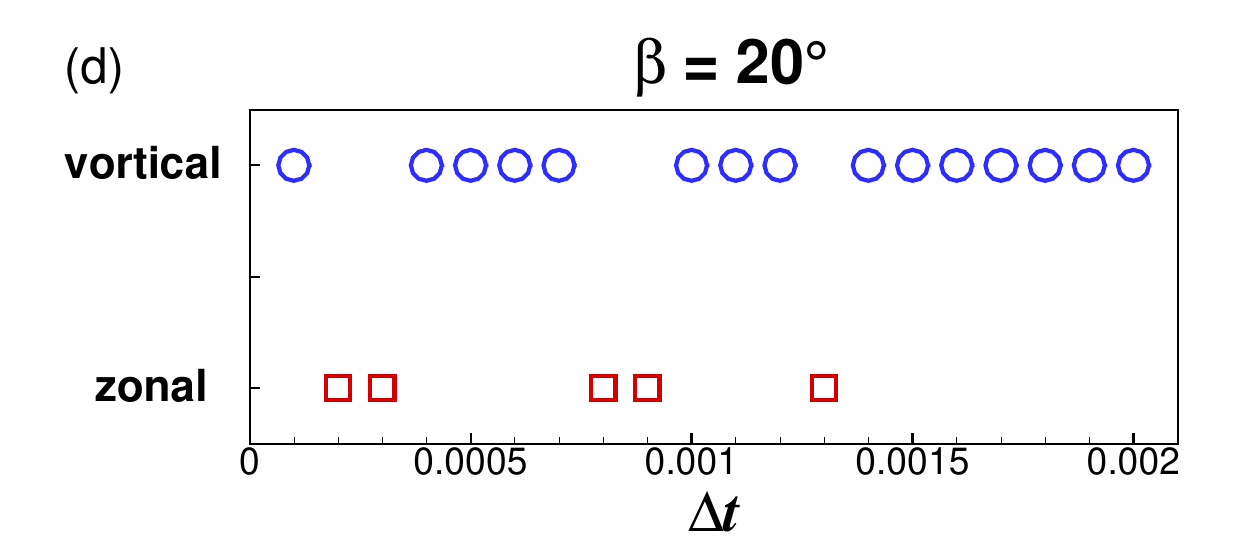}
        \end{tabular}
    \caption{Final flow type   {of tilted} turbulent RBC versus time step $\triangle t$ of DNS: either vortical/roll-like flow (blue circle) or zonal flow (red square). Tilt angle: (a) $\beta=10^\circ$, (b) $\beta=15^\circ$, (c) $\beta=18^\circ$, and (d) $\beta=20^\circ$.}     \label{D10-20}
    \end{center}
\end{figure}

\begin{figure}
    \begin{center}
        \begin{tabular}{cc}
             \includegraphics[width=2.5in]{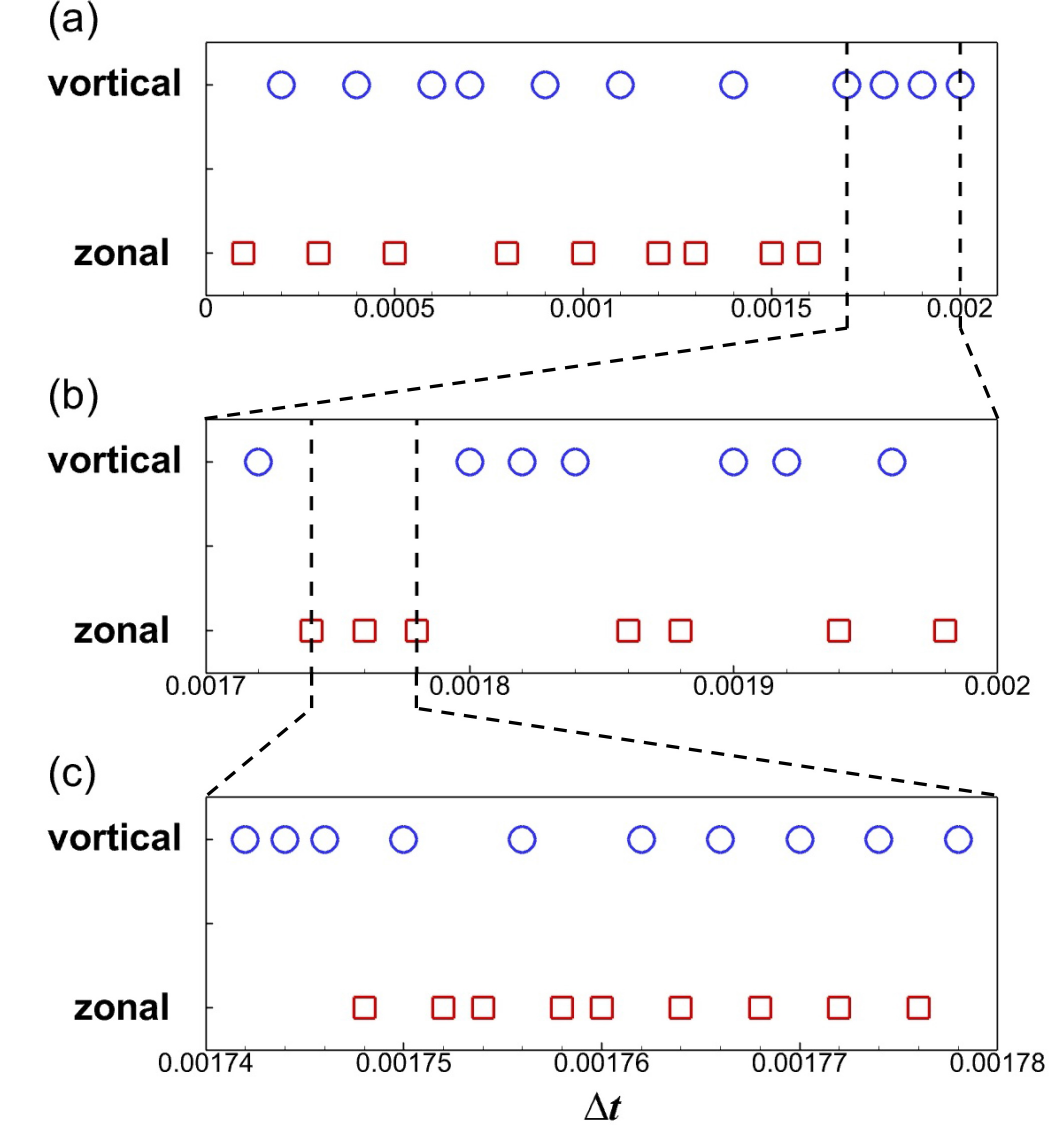}
        \end{tabular}
    \caption{Final flow type   {of turbulent} RBC tilted at the angle $\beta=18^\circ$ versus time step $\triangle t$ of DNS: either vortical/roll-like flow (blue circle) or zonal flow (red square). (a) $\triangle t \in [0.0001, 0.002]$, (b) $\triangle t \in [0.00172, 0.00198]$, and (c) $\triangle t \in [0.001742, 0.001778]$.}     \label{D18-M}
    \end{center}
\end{figure}

\section{Sensitivity dependence of DNS results on time-step}

The Navier-Stokes turbulence described  in Section 1 is   predicted by    traditional direct numerical simulation (DNS)  \cite{Orszag1970, She1990Nature, MoinARFM1998, Scardovelli1999ARFM,Huang2022JFM} using   double-precision arithmetic for all data,  a fourth-order Runge-Kutta method   with time-step $\Delta t  \in [0.0001, 0.002]$ for time integration, and   {a} Fourier pseudo-spectral method with the $3/2$ rule for dealiasing  on the same uniform mesh $N_x\times N_z =1024\times1024$.     
The corresponding spatial resolution is sufficiently fine that the horizontal (maximum) grid spacing is less than the Kolmogorov scale, as previously verified by Pope~\cite{pope2001turbulent}.    In all of our simulations, the Courant-Friedrichs-Lewy (CFL) condition, i.e., Courant number $<1$,  is satisfied by suitable selection of the time-step $\Delta t$.  Therefore, from the traditional viewpoint of DNS, all of our DNS results are reliable and  thus can be regarded as ``benchmark solutions''.  

Without loss of generality, FIG.~\ref{Contour} shows the temperature departure fields  $\theta$ at $t=500$ (i.e., sufficiently long for a stable flow to be attained) for the 2D turbulent RBC tilted at angles $\beta=10^\circ$, $15^\circ$, $18^\circ$, and $20^\circ$, given by DNS  using two different time-steps $\Delta t=7\times10^{-4}$ and  $8\times10^{-4}$.   For  $\beta=10^\circ$, $15^\circ$, $18^\circ$, and $20^\circ$, two completely different types of turbulent  flow  occur , i.e.,  vortical/roll-like flow (left of FIG.~\ref{Contour}) and zonal flow  (right of FIG.~\ref{Contour}).  

Furthermore, to investigate the sensitivity dependence of the final flow type of the 2D turbulent RBC on the time-step $\Delta t$, we   {examined the use of different time-steps} $\Delta t\in [0.0001, 0.002]$  with increment 0.0001  for $\beta=10^\circ$, $15^\circ$, $18^\circ$, and $20^\circ$.    FIG.~\ref{D10-20} shows that the vortical and zonal flow states alternate  randomly   with $\Delta t$.   

Such kind of randomness is more evident through use of a smaller time-step increment.   Without loss of generality, let us consider the case   {when} $\beta = 18^\circ$.  As shown in FIG.~\ref{D10-20}(c) and FIG.~\ref{D18-M}(a),  the DNS results  
given by  $\Delta t = 0.0017$, 0.0018, 0.0019, and 0.002   {with time-step} increment $10^{-4}$ correspond to the {\em same} type of turbulence,  i.e., vortical flow.  However, using a smaller time-step increment $2\times 10^{-5}$ within $\Delta t \in [0.00172, 0.00198]$,  we find that both the vortical and zonal flow types again appear randomly in our DNS results, as shown in FIG.~\ref{D18-M}(b).   Similarly,  only the zonal flow appears within $\Delta t \in [0.00174, 0.00178]$, as shown in 
 FIG.~\ref{D18-M}(b). However, using a even smaller time-step increment $2\times 10^{-6}$,  both of the vortical and zonal flow types again appear randomly, as shown in FIG.~\ref{D18-M}(c).  Clearly,  FIG.~\ref{D18-M}  suggests that the time-step corresponding to each type   {of turbulent flow} should be densely distributed.   Thus,  the final flow type of the 2D turbulent RBC is rather sensitive to the time-step $\Delta t$.

It should be emphasized that, except for the time-step $\Delta t$, {\em all} other aspects such as the initial condition, the uniform mesh, the adapted numerical algorithms and so on,  have remained the {\em same}.  Thus, the choice of time-step $\Delta t$  is  the only  explanation for the qualitative differences between the final flow types.  Therefore, given that the two different values of the time-step $\Delta t$ correspond to two different numerical  forms of noise, the final  characteristics of numerical simulation of the NS turbulence are extremely sensitive to numerical noise that can be regarded as tiny artificial stochastic disturbances  for $\beta=10^\circ$, $15^\circ$, $18^\circ$, and $20^\circ$.  Note that the statistics of different types of numerical simulation of the NS turbulence are entirely different.  Thus,  numerical noise  can  have a huge impact even  on the statistics  of numerical simulation of NS turbulence, such as for the cases of turbulent Rayleigh-B\'{e}nard convection considered in this paper.   

\section{ Discussion and conclusions} 

It is widely accepted by the community of turbulence modellers that DNS results are reliable  provided the following two conditions are satisfied: 
 \begin{enumerate}
\item the horizontal (maximum) grid spacing is less than the Kolmogorov scale;
\item   {the Courant-Friedrichs-Lewy (CFL)} condition   { is satisfied, i.e., the} Courant number $<1$. 
\end{enumerate}
 On carrying out a thorough check, we found that all the DNS results reported herein satisfy both the aforementioned conditions.  Thus, from the traditional viewpoint of DNS, all of our DNS results  should be reliable and could  therefore be used as   benchmark solutions.  However, the final type  of NS turbulence produced is sensitive to the time-step $\Delta t$, which unfortunately is  a numerical factor that is artificially prescribed by the modeller.   So, which DNS result should we believe?  Obviously, turbulent flow is a kind of natural  phenomenon  and certainly should {\em not} be  subject to the free choice of the modeller.  This implies that something is amiss here.  We will reveal the reason below.   

 The cases of turbulent Rayleigh-B\'{e}nard convection considered in this paper  produce a kind of NS turbulence   that is governed by the {\em deterministic}  governing equations (\ref{RB_psi})-(\ref{periodic}), which neglect all small stochastic disturbances for $t>0$, no matter whether they are physical or artificial. It is a pity that  the nonlinear partial differential equations   which comprise the NS equations have no  analytical solution  but instead necessitate numerical treatment.  In practice, all numerical methods are   unavoidably contaminated by numerical noise from truncation and round-off errors.   For a non-chaotic system, numerical   {noise does not amplify as the computation progresses and thus numerical simulation remains negligible.  However, for a chaotic system,   numerical noise can have a huge influence on the results even regarding their statistics, as reported by Lorenz~\cite{Lorenz2006Tellus}.   It is well established that NS turbulence is inherently chaotic ~\cite{Deissler1986PoF, boffetta2017chaos, berera2018chaotic, Vassilicos2023JFM, Ge2025JFM, Ge2025PRF, Okamoto2025PRF}.  Thus, due to the famous butterfly-effect of chaos, the unavoidable  numerical noise in DNS exponentially  increases to a macro-level, which can  then alter the type of the final flow (and  its  statistics).  
 
To clearly reveal it, Liao \cite{liao2026arXivKolmogorovProof} currently proved such a mathematical theorem about the spatial symmetry invariance of incompressible Navier-Stokes equations under periodic boundary condition that, if the initial condition of NS equations has a kind of spatial symmetry,  its solution should remain the {\em same} spatial symmetry {\em forever}, i.e. for all $t\geq 0$.   However, as illustrated by Qin {\em et al.} \cite{Qin2024JOES, Qin2026EJMB, Qin2025PoF},  the DNS results of turbulent Kolmogorov flow under an initial condition with spatial symmetry quickly lose the spatial symmetry and thus violate the above-mentioned mathematical theorem: this clearly indicates that the DNS results are indeed quickly polluted by numerical noises badly.   They are very good examples~\cite{Qin2024JOES, Qin2026EJMB, Qin2025PoF, Liao-2025-JFM-NEC, Qin2022JFM}  to illustrate that  numerical noise of DNS results indeed might have significant influences  on macro-scale property of turbulence such as spatial symmetry, statistics and so on.      
 
All of these indicates that  numerical noise as a small stochastic disturbance  exerts a huge influence on NS turbulence.  Unfortunately, NS turbulence as a {\em deterministic} model neglects {\em all} stochastic disturbances for $t>0$.  In logic, this leads to a serious  {\em paradox}  for NS turbulence as a mathematical model.  Note that NS equations were first proposed  by Navier~\cite{Navier1822} in 1822 and then modified by Stokes~\cite{Stokes1845} in 1845, many years  before  Poincar\'{e}~\cite{Poincare1890} proposed, for the first time,  the concept of sensitivity dependence on initial condition  in 1890, and more than one century before the chaos theory~\cite{Lorenz1963, Smale1967, Li-Yorke1975, May1976, Feigenbaum1978, Feigenbaum1979} was well established, and even many years before  the concept of turbulence was proposed by Reynolds~\cite{Reynolds1895} in 1895 and by Taylor~\cite{Taylor1935} in 1935.   So, it was very reasonable for Navier~\cite{Navier1822}  and Stokes~\cite{Stokes1845} to completely neglect the influence of all tiny disturbances in the Navier-Stokes equations.   However, this leads to a paradox in logic, and we had to face its challenge today.    

It should be emphasized  that NS turbulence is simply a mathematical model.  It is {\em not} the  actual turbulence  encountered in nature.  In fact, the {\em exact}  solution of NS turbulence is different from the DNS result  of NS   turbulence -- the latter is badly contaminated by numerical noise,  but the former is {\em not}.   Therefore,  the following three  forms of turbulence, i.e.,
  \begin{enumerate}
\item[(A)] exact solution of NS turbulence,
\item[(B)] numerical simulation of NS turbulence,  and
\item[(C)] real turbulence,
\end{enumerate}
might be different.      
 The question as to whether or not {\em numerical} simulations of NS turbulence agree with  the {\em exact} solution (even   in terms of statistics) and can describe correctly  the essential characteristics of the corresponding {\em  actual} turbulence   implies a need for substantial further theoretical, numerical, and experimental investigations.  Although many numerical and experimental results support the use of NS turbulence as a satisfactory model, the DNS results reported in our paper reveal a  logical paradox affecting the theory  underpinning NS turbulence. We strongly recommend that small physical/artificial  stochastic disturbances should be considered in turbulence modelling.  In practice, tiny physical disturbances should be incorporated in the Navier-Stokes equations for $t>0$, just like the Landau-Lifshitz-Navier-Stokes equations \cite{LLNS1959}  which are stochastic partial differential equations. 

\vspace{0.5cm}
	
\hspace{-0.65cm}{\bf Acknowledgments} This work is partly supported by the National Natural Science Foundation of China (No. 12302288 and No. 12521002),  the State Key Laboratory of Ocean Engineering, and the Dept. of Mathematics, Hong Kong University of Science and Technology.

\section*{Reference}

\bibliography{ultrachaos-BIB}
	
\end{document}